\documentclass[12pt]{iopart}

\expandafter\let\csname equation*\endcsname\relax
\expandafter\let\csname endequation*\endcsname\relax
\usepackage{amsmath}
\UseRawInputEncoding
\usepackage{graphicx}
\usepackage{subfigure}
\usepackage[ruled, linesnumbered]{algorithm2e}
\usepackage{setspace}
\usepackage{array}
\usepackage{booktabs}
\usepackage{tabularx}
\usepackage{multirow}
\usepackage{hyperref}
\usepackage{cleveref}
\hypersetup{hypertex=true,
	colorlinks=true,
	linkcolor=blue,
	anchorcolor=blue,
	citecolor=blue}
\usepackage{graphbox}
\usepackage{appendix}

\begin{document}

\title[]{Material decomposition for dual-energy propagation-based phase-contrast CT}

\author{Suyu Liao$^1$, Huitao Zhang$^1$, Peng Zhang$^1$ and Yining Zhu$^{1}$}

\address{$^1$ School of Mathematical Sciences, Capital Normal University, Beijing, 100048, China}

\ead{suyuliao6@gmail.com}

\vspace{10pt}
\begin{indented}
\item[Received] xxxxxx
\item[Accepted for publication] xxxxxx
\item[Published] xxxxxx
\end{indented}

\begin{abstract}
Material decomposition refers to using the energy dependence of material physical properties to differentiate materials in a sample, which is a very important application in computed tomography(CT). In propagation-based X-ray phase-contrast CT, the phase retrieval and Reconstruction are always independent. Moreover, like in conventional CT, the material decomposition methods in this technique can be classified into two types based on pre-reconstruction and post-reconstruction (two-step). The CT images often suffer from noise and artifacts in those methods because of no feedback and correction from the intensity data. This work investigates an iterative method to obtain material decomposition directly from the intensity data in different energies, which means that we perform phase retrieval, reconstruction and material decomposition in a one step. Fresnel diffraction is applied to forward propagation and CT images interact with this intensity data throughout the iterative process. Experiments results demonstrate that compared with two-step methods, the proposed method is superior in accurate material decomposition and 
noise reduction.

\vspace{1pc}
\end{abstract}

\section{Introduction}
\label{sect1}
Conventional computed tomography (CT) has been widely used in medical diagnosis and industrial testing. Since the principle is based on the variability of attenuation between different substances, it is hard to perform visualization for the weakly difference in attenuation between two materials (such as brain, breast, lung, etc.). Fortunately, phase-contrast imaging that is sensitive to the phase-shifting properties of the object has been introduced. The differences in X-ray phase shifts in low-z materials are about three orders of magnitude larger than their absorption changes \cite{Richard2000,Atsushi1995}, which is great beneficial for high spatial resolution of sample in pre-clinical studies. There are some techniques that can measure the phase-shifting properties of the objects, such as analyzer-based \cite{Zhang2010, Luigi2007}, grating-based \cite{2006Phase,Wang2014}, edge-illumination \cite{PMID:22891301,Olivo_2021}, and propagation-based imaging \cite{Snigirev1995On,1996Phase}. In this paper, we focus on propagation-based imaging because it does not require other optical components to facilitate the implementation.

Material decomposition is an important application in CT, which achieves quantitative material images and artifact reduction. The existing methods for material decomposition in dual-energy conventional CT can be classified into several groups: image-based methods, projection-based methods, iterative methods. 
In image-based method, the projection data sets in different spectra are independent and material decomposition is carried out after tomographic reconstruction \cite{Johns1985Theoretical,7239569}. The decomposition results always suffer from artifacts because of failure for describing the real non linearity between the projection and decomposed results. The projection-based methods refer to getting firstly basic material projection and then reconstruction. Compared with image-based method, the projection-based method can obtain better decomposition results, however, the projections must satisfy the requirement of geometric consistency  \cite{2006First,2007Empirical}. Iterative methods based on statistical models and nonlinear optimizations are proposed, which effectively improve the image quality by incorporating prior information or establishing approximate model \cite{993128,10.1117/12.812391,2014Multi,6965595,CHEN2021101821}. Those iterative methods perform tomographic reconstruction and material decomposition in a one-step. 

In recent years, X-ray phase-contrast imaging has made great progress for material quantitation and decomposition, especially for grating-based imaging \cite{9078041,2020A}. Using attenuation, refraction and dark-filed information of the sample, accurate three-material decomposition from dual-energy differential phase-contrast CT was presented \cite{2020Single}. Mechlem et al. confirmed that spectral grating-based phase-contrast imaging could strongly reduce the noise level of the image \cite{8784289}.
Although obtaining good decomposition results, grating-based imaging based on the grating self-imaging effect must use a well-built gratings, which makes the experimental operation complicated and reduces X-ray flux to the sample.
Fortunately, the propagation-based imaging can effectively avoid the above problem. Similar work has previously been achieved for material quantitation and decomposition using this technique \cite{GUREYEV2010Quantitative,2001Quantitative,Ghani:21,liao}. Li et al. proposed an iterative method for solving the phase-retrieval problem by Alvarez-Macovski model, then obtain photoelectric absorption image and electron density image, and finally material decomposition \cite{Li_2020}. Schaff et al. investigated an analytical method in propagation-based phase-contrast imaging to perform phase retrieval and material decomposition in one step \cite{9133129}. We know that the propagation-based imaging can incorporate CT to obtain the tomography of the weakly attenuation object. But it is also carried out in two steps: step (1): the projection of phase retrieval or material decomposition should be calculated; step (2): tomographic reconstruction by Filter Back Projection (FBP) or Algebraical Reconstruction Technique(ART) \cite{GORDON1970471}. The tomographic image, however, usually suffer from noise and artifacts since step (1) and step (2) are independent without feedback from the intensity data.

In this paper, inspired by Schaff's work and iterative methods in conventional CT, we propose a one-step approach to perform material decomposition directly from intensity data in dual-energy propagation-based phase-contrast CT (OMD-PPCT), which means that phase retrieval, tomographic reconstruction, and material decomposition are performed in the one step. In addition, we choose Fresnel wave propagation to simulate the forward propagation in this iterative process, which is more consistent with the physical mechanism at low energy. Moreover, we choose the Simultaneous Algebraic Reconstruction Technique (SART) instead of the Algebraic Reconstruction Technique in this iterative to speed up the convergence and reduce the time consumption \cite{1984Simultaneous}. The tomographic images of the proposed method are compared with that of previous methods in Poisson noise case.

The remainder of this paper is organized as follows. In section 2, we introduce the imaging model and propose an algorithm for material decomposition. In section 3, numerical experiments are provided to verify the proposed algorithm. Further discussions are shown in section 4 and we conclude the paper in section 5.

\section{Method}

\subsection{Imaging model}
In conventional CT, the basic imaging model can be described by Lambert-Beer Law:
\begin{equation}
\label{equ:1}
{I(x)} = {I^{in}}\exp ( - \int {\mu (x,y) dy} )
\end{equation}
here $\mu$ is the linear attenuation coefficient. In fact, the interaction between sample with X-ray also can be described by the complex refractive index:
\begin{equation}
n = 1 - \delta  + {\bf{i}}\beta
\end{equation}
where $\delta$ is the real decay rate of the refractive index, and $\mu  = \frac{{4\pi }}{\lambda }\beta$, $\lambda$ the wavelength. When X-ray plane wave pass through the object, the wave
function of the emergent beam reads:
\begin{equation}
A(x) = {A^{in}}\exp ( - \frac{{M(x)}}{2} + {\bf{i}}\Phi (x))
\end{equation}
the $A^{in}$ is incident X-ray plane wave, $\Phi (x) = - \frac{{2\pi }}{\lambda }\int {\delta (x,y)dy} $, and $M(x) = \frac{{4\pi }}{\lambda }\int {\beta (x,y)dy} $. Assuming the object is placed at a distance $z$ from the detector (ODD) . $I_z$ the intensity at a distance $z$, when $z=0$, the intensity is
\begin{equation}
I_0(x) = {\left| {A(x)} \right|^2} = {I^{in}}\exp ( - M(x))
\end{equation}
Obviously, the above function is also the eq.\ref{equ:1}. the plane intensity based on Fresnel diffraction theory at the distance $z$ can be written as
\begin{equation}
\label{equ:decom5}
{I_z}(x) = {\left| {{h_z} \otimes A(x)} \right|^2}
\end{equation}
The Fresnel propagator is ${h_z} = \frac{{\exp ({\bf{i}}kz)}}{{{\bf{i}}\lambda z}}\exp ({\bf{i}}\frac{\pi }{{\lambda z}}({x^2}))$, here $k = \frac{{2\pi }}{\lambda }$. Note that when $I_z$ is 2-Dimensional array, the ${h_z} = \frac{{\exp ({\bf{i}}kz)}}{{{\bf{i}}\lambda z}}\exp ({\bf{i}}\frac{\pi }{{\lambda z}}({x^2}+{y^2}))$.

The transport of intensity equation (TIE) can describe the X-ray wave for propagation in z \cite{GUREYEV2010Quantitative}:
\begin{equation}
{I_z(x)} = {I_0(x)}(1 - \frac{{z\lambda }}{{2\pi }}{\nabla ^2}\Phi (x)) - \frac{{z\lambda }}{{2\pi }}\nabla {I_0(x)}\nabla \Phi(x)
\end{equation}
Assuming that there are no large intensity gradients in $I_0$, we can ignore the last term \cite{Dpaga} and get:
\begin{equation}
{I_z(x)} = {I_0(x)}(1 - \frac{{z\lambda }}{{2\pi }}{\nabla ^2}\Phi(x) )
\end{equation}
assume $\frac{{z\lambda }}{{2\pi }}{\nabla ^2}\Phi(x)  \ll 1$, the above equation can be simplified:
\begin{equation}
{I_z(x)} = {I_0(x)}\exp (- \frac{{z\lambda }}{{2\pi }}{\nabla ^2}\Phi(x) )
\end{equation}
As a last step, we take the logarithm and arrive at the linearized TIE:
\begin{equation}
- \ln [{I_z}(x)] = \int {\mu (x,y)dy - z{\nabla ^2}} \int {\delta (x,y)dy}
\end{equation}

\subsection{Algorithm} 
In this part, we will derive a new method that can obtain directly the basic material image from the original data without phase retrieval in the process. For two or more basic materials of the sample, the attenuation and phase shift can be defined by a set of basic functions. Here we focus on two different material decomposition, which can be carried out:
\begin{equation}
\left\{ {\begin{array}{*{20}{c}}
	{\mu (x,y) = f(x,y){\mu _1} + g(x,y){\mu _2}}\\
	{\delta (x,y) = f(x,y){\delta _1} + g(x,y){\delta _2}}
	\end{array}} \right.
\end{equation}
Adding the basic material information, the linearized TIE can be transformed into
\begin{equation}
	\begin{aligned}
			- {\rm{\ln}}[I({x, E_i})] &= {\mu _{1,i}}\int {f(x,y)dy} + {\mu _{2,i}}\int {g(x,y)dy}\\
			&- z{\nabla ^2}[{\delta _{1,i}}\int {f(x,y)dy} + {\delta _{2,i}}\int {g(x,y)dy} ]\\
			&=({\mu _{1,i}} - z{\nabla ^2}{\delta _{1,i})}\int {f(x,y)dy}  + {\rm{(}}{\mu _{2,i}}
			- z{\nabla ^2}{\delta _{2,i}})\int {g(x,y)dy}
	\end{aligned}
\end{equation}
here $i = 1,2$, $I(E_1)$ and $I(E_2)$ are the intensities at different energy. Then organizing the above equation, we can obtain
\begin{equation}
- \mathcal{F}({\rm{\ln }}I({E_i})){\rm{  =  (}}{\mu _{1,i}} - z{{\bf{\tau }}^2}{\delta _{1,i}})\mathcal{F}(\int {f(x,y)dy)}  + {\rm{(}}{\mu _{2,i}} - z{{\bf{\tau }}^2}{\delta _{2,i}})\mathcal{F}(\int {g(x,y)dy)}
\end{equation}
here $\mathcal{F}$ is the Fourier transform operator. ${\tau ^2}$ is the Fourier components of ${\nabla ^2}$.

Let ${\bf{f}} {\rm{ = (}}{f _1}{\rm{,}}{f _2}, \ldots {f _J}{{\rm{)}}^\tau }$, and ${\bf{g}} {\rm{ = (}}{g _1}{\rm{,}}{g _2}, \ldots {g _J}{{\rm{)}}^\tau }$  denote the discretized images of $f(x,y)$ and $g(x,y)$, where $f_j$ and $g_j$ are the sampled values of $f(x,y)$ and $g(x,y)$ at the $j$th pixel, $J$ the total pixel number, and $\tau$ the vector transpose operation. ${R^\varphi } = {(r_{uj}^\varphi )_{U \times J}}$ is the projection matrix at angle $\varphi$, where $(r_{uj}^\varphi )$ represents the contribution of $f_j$ and $g_j$ to the projection along the $u$-th x-ray path at projection angle $\varphi$. U is the number of detector cell.
We obtain the X-ray intensity in m-iteration, that is
\begin{equation}
- \mathcal{F}({\rm{\ln [}}I({E_i})^{\varphi,m}{]}){\rm{  =  (}}{\mu _{1,i}} - z{{\bf{\tau }}^2}{\delta _{1,i}})\mathcal{F}({R^\varphi }{{\bf{f}}^m}) + {\rm{(}}{\mu _{2,i}} - z{{\bf{\tau }}^2}{\delta _{2,i}})\mathcal{F}({R^\varphi }{{\bf{g}}^m})
\end{equation}
The residual X-ray intensity:
\begin{equation}
\label{equ:decom14}
	\begin{aligned}
{{\bf{I}}_i} &=  - \mathcal{F}({\rm{\ln }}I({E_i})^\varphi) - [ - \mathcal{F}({\rm{\ln }}I{({E_i})^{\varphi ,m}})\\
  &=  ({\mu _{1,i}} - z{{{\tau }}^2}{\delta _{1,i}})\mathcal{F}[{R^\varphi }({\bf{f}} - {\bf{f}^m})]
 + {{(}}{\mu _{2,i}} - z{{{\tau }}^2}{\delta _{2,i}})\mathcal{F}[{R^\varphi }(\bf{g} - {\bf{g}^m})]
\end{aligned}
\end{equation}
let ${f_{err}} = \mathcal{F}({R^\varphi }(\bf{f} - {\bf{f}^m})$, and ${g_{err}} = \mathcal{F}({R^\varphi }(\bf{g} - {\bf{g}^m})$, we get
\begin{equation}
\left\{ {\begin{array}{*{20}{c}}
	{{{\bf{I}}_1}{\rm{ =  (}}{\mu _{1,1}} - z{{\bf{\tau }}^2}{\delta _{1,1}}){f_{err}} + {\rm{(}}{\mu _{2,1}} - z{{\bf{\tau }}^2}{\delta _{2,1}}){g_{err}}}\\
	{{{\bf{I}}_2}{\rm{ =  (}}{\mu _{1,2}} - z{{\bf{\tau }}^2}{\delta _{1,2}}){f_{err}} + {\rm{(}}{\mu _{2,2}} - z{{\bf{\tau }}^2}{\delta _{2,2}}){g_{err}}}
	\end{array}} \right.
\end{equation}
Solving the above equation,
\begin{equation}
\label{equ:decom16}
\left\{ {\begin{array}{*{20}{c}}
	{{f_{err}} = \frac{{{\bf{I}_1}{\rm{(}}{\mu _{2,2}} - z{{\bf{\tau }}^2}{\delta _{2,2}}) - {\bf{I}_2}{\rm{(}}{\mu _{2,1}} - z{{\bf{\tau }}^2}{\delta _{2,1}})}}{{{\rm{(}}{\mu _{2,2}} - z{{\bf{\tau }}^2}{\delta _{2,2}}){\rm{(}}{\mu _{1,1}} - z{{\bf{\tau }}^2}{\delta _{1,1}}) - {\rm{(}}{\mu _{2,1}} - z{{\bf{\tau }}^2}{\delta _{2,1}}){\rm{(}}{\mu _{1,2}} - z{{\bf{\tau }}^2}{\delta _{1,2}})}}}\\
	{{g_{err}} = \frac{{{\bf{I}_1}({\mu _{1,2}} - z{{\bf{\tau }}^2}{\delta _{1,2}}) - {\bf{I}_2}({\mu _{1,1}} - z{{\bf{\tau }}^2}{\delta _{1,1}})}}{{{\rm{(}}{\mu _{2,1}} - z{{\bf{\tau }}^2}{\delta _{2,1}}){\rm{(}}{\mu _{1,2}} - z{{\bf{\tau }}^2}{\delta _{1,2}}) - {\rm{(}}{\mu _{1,1}} - z{{\bf{\tau }}^2}{\delta _{1,1}}){\rm{(}}{\mu _{2,2}} - z{{\bf{\tau }}^2}{\delta _{2,2}})}}}
	\end{array}} \right.
\end{equation}
Then combing the SART algorithm, we can reconstruct the basic images $\bf{f}$ and $\bf{g}$ for $m+1$ iteration, the scheme is as follows:
\begin{equation}
\label{equ:decom17}
\left\{ {\begin{array}{*{20}{c}}
	{{\bf{f}}_j^{m + 1} = {\bf{f}}_j^m + \frac{\varepsilon }{{R_{ + ,j}^\varphi }}\sum\limits_{u = 1}^U {\frac{{r_{u,j}^\varphi }}{{R_{u, + }^\varphi }}{{[{\mathcal{F}^{ - 1}}({f_{err}})]}_u}} }\\
	{{\bf{g}}_j^{m + 1} = {\bf{g}}_j^m + \frac{\varepsilon }{{R_{ + ,j}^\varphi }}\sum\limits_{u = 1}^U {\frac{{r_{u,j}^\varphi }}{{R_{u, + }^\varphi }}{{[{\mathcal{F}^{ - 1}}({g_{err}})]}_u}} }
	\end{array}} \right.
\end{equation}
where $R_{u, + }^\varphi  = \sum\nolimits_{j = 1}^J {r_{u,j}^\varphi }$ with $u = 1,2, \ldots U$, and $R_{ + ,j}^\varphi  = \sum\nolimits_{u = 1}^U {r_{u,j}^\varphi }$ with $j = 1,2, \ldots J$. $\varepsilon$ the relaxation factor. In addition, we choose Fresnel wave propagation to simulate the forward propagation in this iterative process. Since the Fresnel propagator is a Gaussian-like function, the convolution operation indicates it plays a role in spreading and smoothing the wavefront in the evolution process. With feedback and correction in this method, the noise can be effectively restrained

We summarize the implementation steps of the iteration scheme as follows.

1: Initialization ${{\bf{f}}^{m}}=0$, ${{\bf{g}}^{m}}=0$, $m=0$ and maximum number of iteration;

2: Use Eq.\ref{equ:decom5} based on Fresnel diffraction theory to obtain $I{({E_i})^{\varphi ,m}}$;

3: Calculate the residual intensity by Eq.\ref{equ:decom14}; Then calculate residual results $f_{err}$ and $g_{err}$ using Eq.\ref{equ:decom16};

4: Iteratively update ${\bf{f}}^{m+1}$ and ${\bf{g}}^{m+1}$ according to Eq.\ref{equ:decom17};

5: Set $m=m+1$ and turn to step (2) until the stop condition is met;

6: Return ${\bf{f}}^{m}$ and ${\bf{g}}^{m}$.

\section{Experiments}

In this section, the proposed algorithm is evaluated by numerical experiments. As a comparison, conventional CT method and material decomposition using spectral propagation-based phase imaging (SPBI) \cite{9133129} are chosen. In fairness, the comparison algorithms use SART for tomographic reconstruction. The conventional CT method we chose is the image-based method, which means that the projection data sets in different energy are independent and material decomposition is carried out after tomographic reconstruction. 

\subsection{Spherical phantom studies}
\begin{figure}[h]
	\centering\includegraphics[scale=0.7]{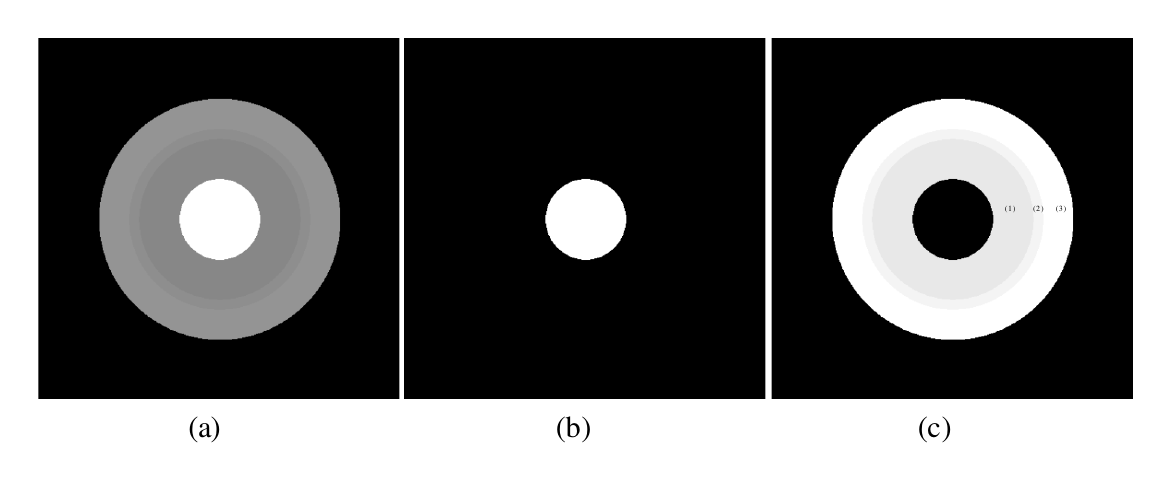}
	\caption{Phantom utilized in the numerical experiments. (a) Numerical phantom; (b) The Teflon component; (c) The Polyme Metha component.}
	\label{Fig.5}
\end{figure}

In this subsection, we designed a sample with two low-Z basic materials. As shown in Fig. \ref{Fig.5}, this sample consists of three spherical layers of concentric homogeneous material with different densities wrapped around a core, namely Teflon and Poly(methyl methacrylate) (PMMA). In Fig. \ref{Fig.5}(c), we assume that PMMA(1) is the standard density of PMMA, PMMA(2)=1.05*PMMA(1) and PMMA(3)=1.1*PMMA(1). The diameter size of the sample is 1.7mm and the experimental parameters are displayed in Table \ref{Tab.4}. A parallel-beam setting was used for acquiring 360 projections equally spaced in 180 degrees in the simulation. Both noise-free and Poisson noise corresponding to emission flux of $10^6$ photon were tested. For material decomposition, the Teflon and PM were chosen as basic material in the multi-material object. The size of the reconstruction was 512*512.

\begin{table}[h]
	\caption{Experimental parameters in numerical simulation.}
	\begin{indented}
		\item[]\begin{tabular}{cc}
			\br
			\mr
			\verb" Energy 1"&15 keV\\
			\verb" Energy 2"&30 keV\\
			\verb" ODD "&20 cm\\
			\verb" Angle Range "&[0,180$^\circ$]\\
			\verb" Projections "&360\\
			\verb" Pixel Array Detector "&512*1\\
			\verb" Detector Unit Size"&5 um\\
			\br
		\end{tabular}
		\label{Tab.4}
	\end{indented}
\end{table}

\begin{table}[t]
	\caption{PSNR and SSIM comparison in noise free case.}
	
	\begin{indented}
		\item[]\begin{tabular}{cccccccc}
			\br
			&&{PM}&&&&{Teflon}\\
			\mr
			&Image-Based&SPBI&OMD-PPCT&&Image-Based&SPBI&OMD-PPCT\\
			\mr
			PSNR&12.133&25.879&$\bf{36.640}$&&34.000&33.000&$\bf{39.922}$\\
			
			SSIM&0.871&0.994&$\bf{0.999}$&&0.994&0.993&$\bf{0.998}$\\
			\br
		\end{tabular}
		\label{Tab.5}
	\end{indented}
\end{table}

\begin{table}[t]
	\caption{PSNR and SSIM comparison in noisy case.}
	\begin{indented}
		\item[]\begin{tabular}{cccccccc}
			\br
			&&{PM}&&&&{Teflon}\\
			\mr
			&Image-Based&SPBI&OMD-PPCT&&Image-Based&SPBI&OMD-PPCT\\
			\mr
			PSNR&6.936&15.886&$\bf{27.211}$&&16.960&21.067&$\bf{32.350}$\\
			
			SSIM&0.667&0.939&$\bf{0.995}$&&0.788&0.905&$\bf{0.992}$\\
			\br
		\end{tabular}
		\label{Tab.6}
	\end{indented}
\end{table}
\begin{figure}[]
	\centering\includegraphics[scale=0.6]{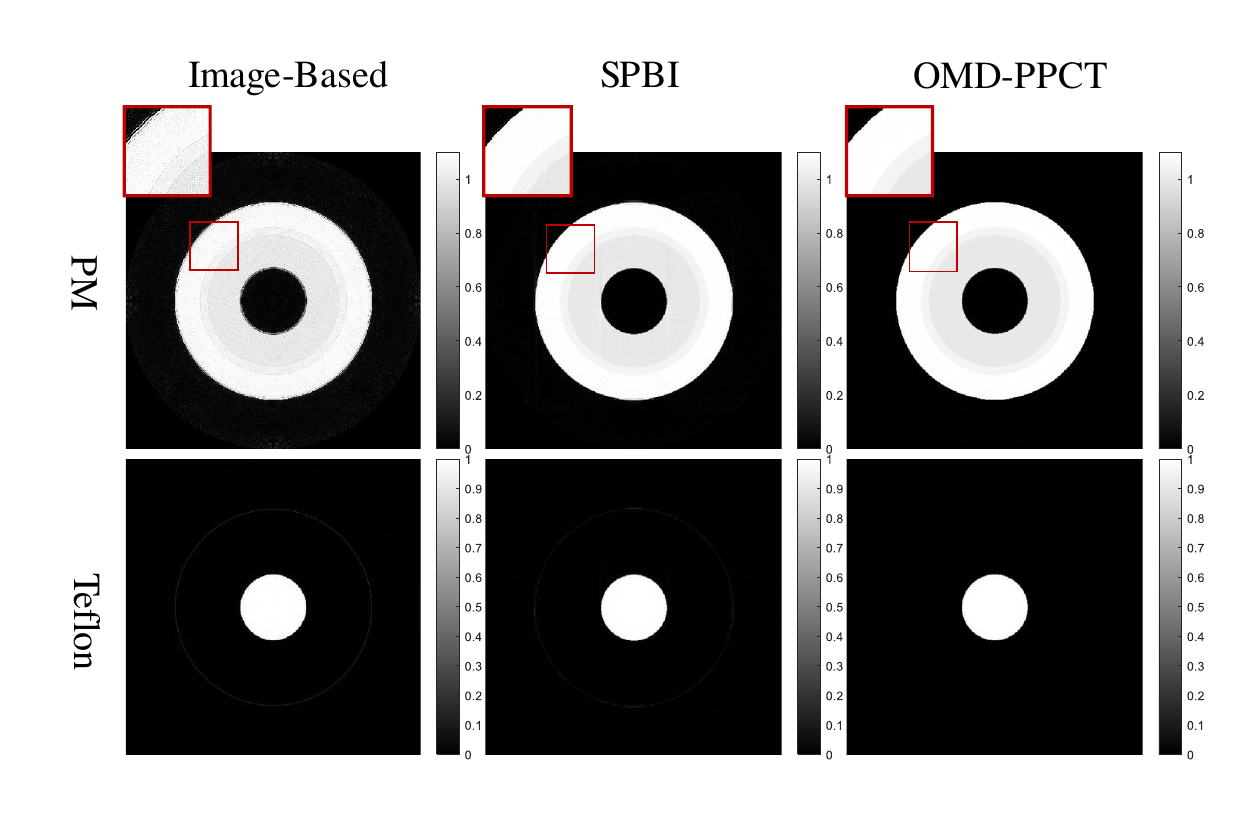}
	\caption{Material decomposition in noise-free case}
	\label{Fig.6}
\end{figure}
\begin{figure}[]
	\centering\includegraphics[scale=0.6]{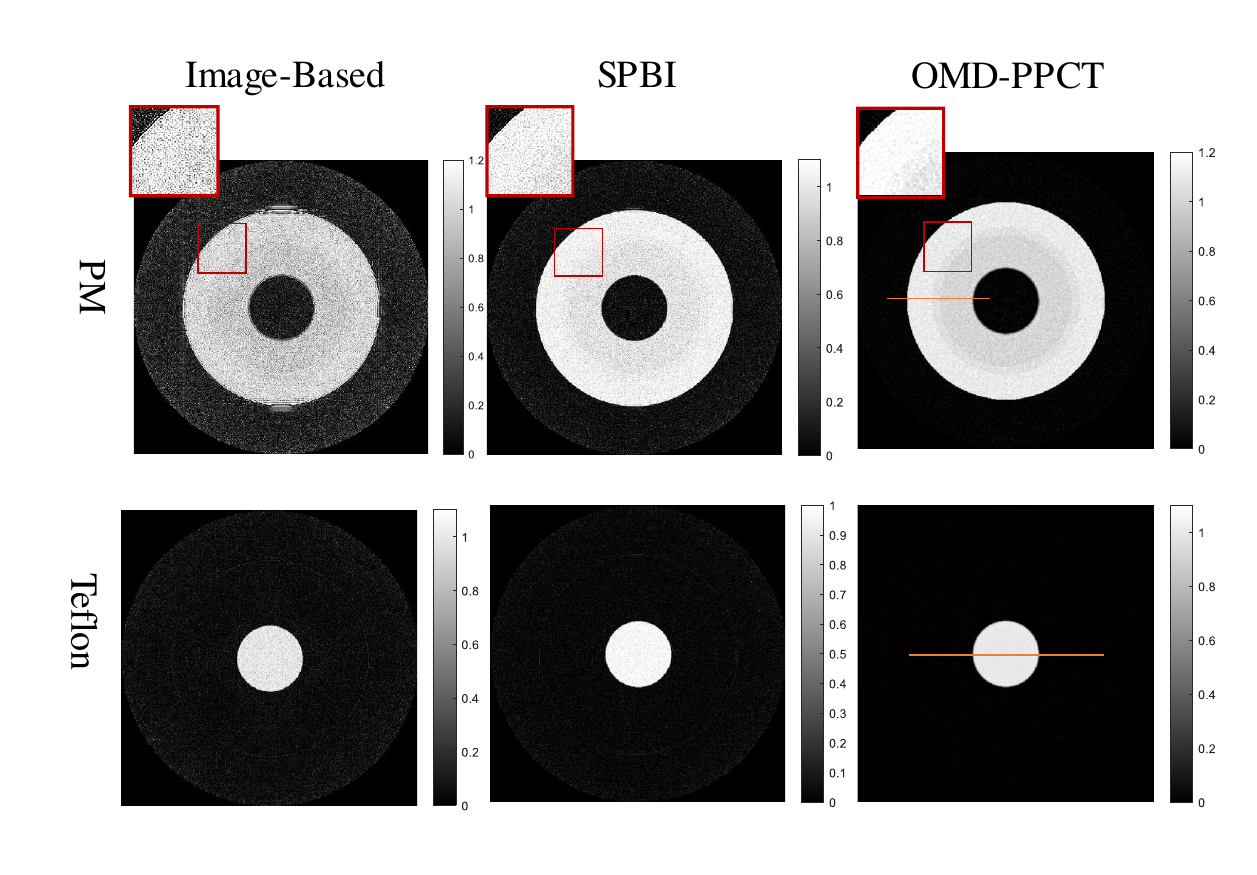}
	\caption{Material decomposition in noisy case}
	\label{Fig.7}
\end{figure}
\begin{figure}[]
	\centering\includegraphics[scale=0.6]{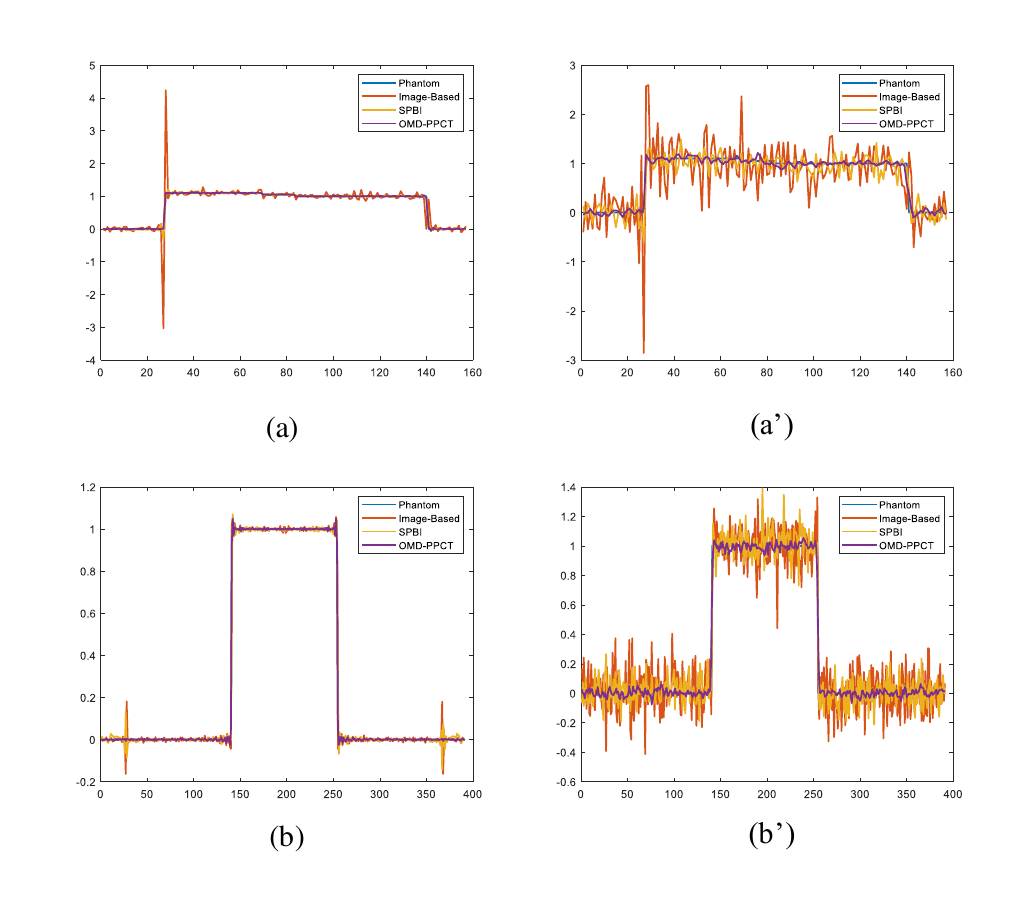}
	\caption{Profiles of decomposed results. (a) and (a') are the PM material results in noise-free and noisy case, respectively; (b) and (b') are the Teflon material results in noise-free and noisy case, respectively. }
	\label{Fig.8}
\end{figure}
The results of basic material decomposition are shown in Fig. \ref{Fig.6} and Fig. \ref{Fig.7}. We also enlarged the contents of the red rectangle. Fig. \ref{Fig.8} shows the profiles of the orange lines of Fig. \ref{Fig.7}. From the profiles and enlarged area, we find that image-based method and SPBI have very weak spatial resolution for PMMA-2, and the proposed algorithm has superior material decomposition and noise suppression than the other methods. It is because that the phase information is not taken into account in image-based method, so there are bright and dark at the edges of water images. Additionally, the SPBI and image-based methods are two-step methods without feedback from the intensity data, which make noise amplification.
The Table \ref{Tab.5} and Table \ref{Tab.6} show Peak Signal to Noise Ratio (PSNR) and Structural Similarity (SSIM) of material decomposition in noise free and noisy cases. These results provide important proof for the ability of accurate material decomposition in OMD-PPCT.

\subsection{biological phantom studies}
\begin{figure}[]
	\centering\includegraphics[scale=0.8]{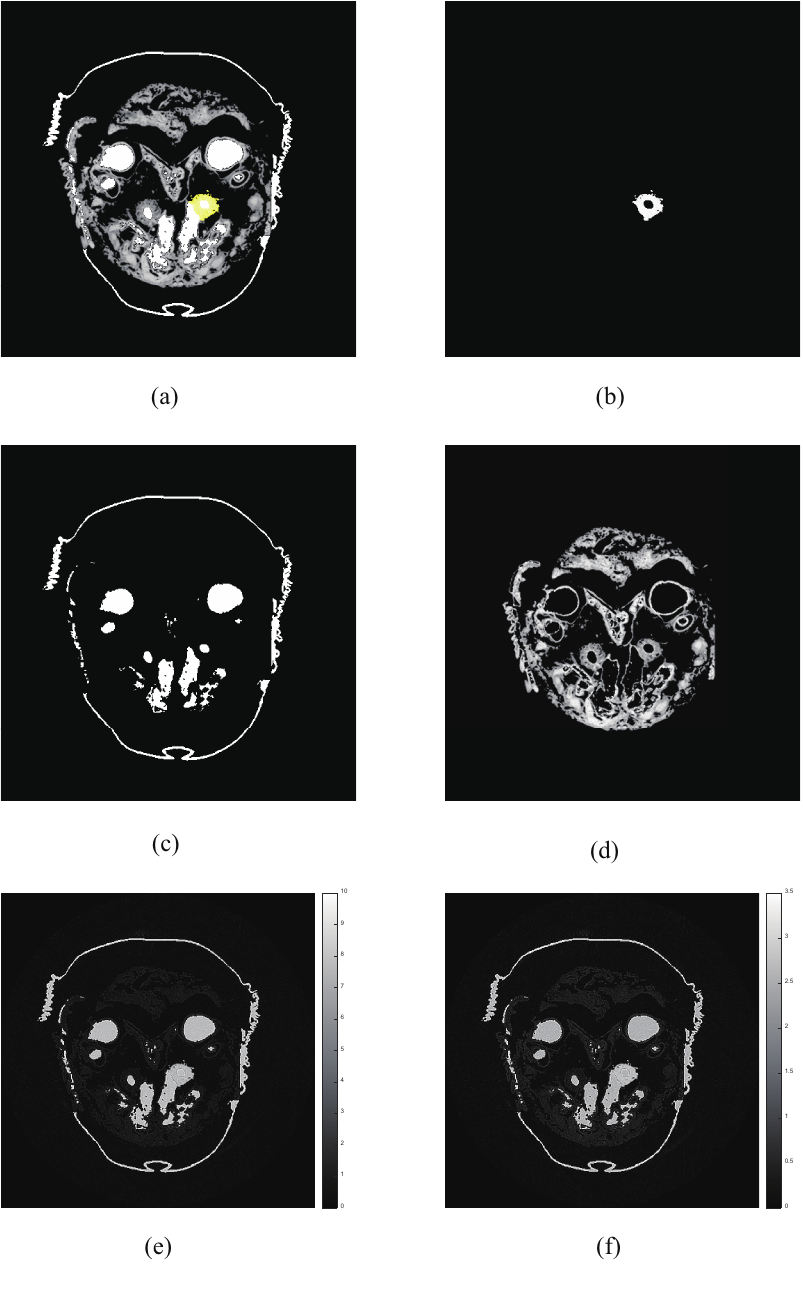}
	\caption{Phantom utilized in the numerical experiments. (a) Numerical phantom; (b) The gold nanoparticles; (c) The bone component; (d) The water component; (e) Conventional CT reconstruction in 20 keV; (f) Conventional CT reconstruction in 30 keV.}
	\label{Fig.1}
\end{figure}
\begin{table}[t]
\caption{Experimental parameters in numerical simulation.}
\begin{indented}
	\item[]\begin{tabular}{cc}
		\br
		\mr
		\verb" Energy 1"&20 keV\\
		\verb" Energy 2"&30 keV\\
		\verb" ODD "&20 cm\\
		\verb" Angle Range "&[0,180$^\circ$]\\
		\verb" Projections "&360\\
		\verb" Pixel Array Detector "&512*1\\
		\verb" Detector Unit Size"&5 um\\
		\br
	\end{tabular}
 \label{Tab.1}
\end{indented}
\end{table}
In medical imaging, nanogold labeling technology has been widely used in the field of immunoassay.  As shown in Fig. \ref{Fig.1}, we designed a pseudo-biological sample with the size of 1.9mm*1.8mm, whose main components are water and bone. The solute with $0.43\%$ concentration of gold nanoparticles are injected in the local area of the water-based material. In this case, it is difficult for conventional CT to distinguish between bone-based material and bone-based material in the gold nanoparticles region because of the similar attenuation values in 20 keV. The experimental parameters are displayed in Table \ref{Tab.1}. Here we also chose parallel-beam setting for 360 projections equally spaced in $180^0$. Both noise-free and Poisson noise cases corresponding to emission flux of $4*10^5$ photons per measurements are tested. Bone and water as the basic materials are selected for material quantitation. The size of the reconstructed image was 512*512.
\begin{figure}[]
	\centering\includegraphics[scale=0.7]{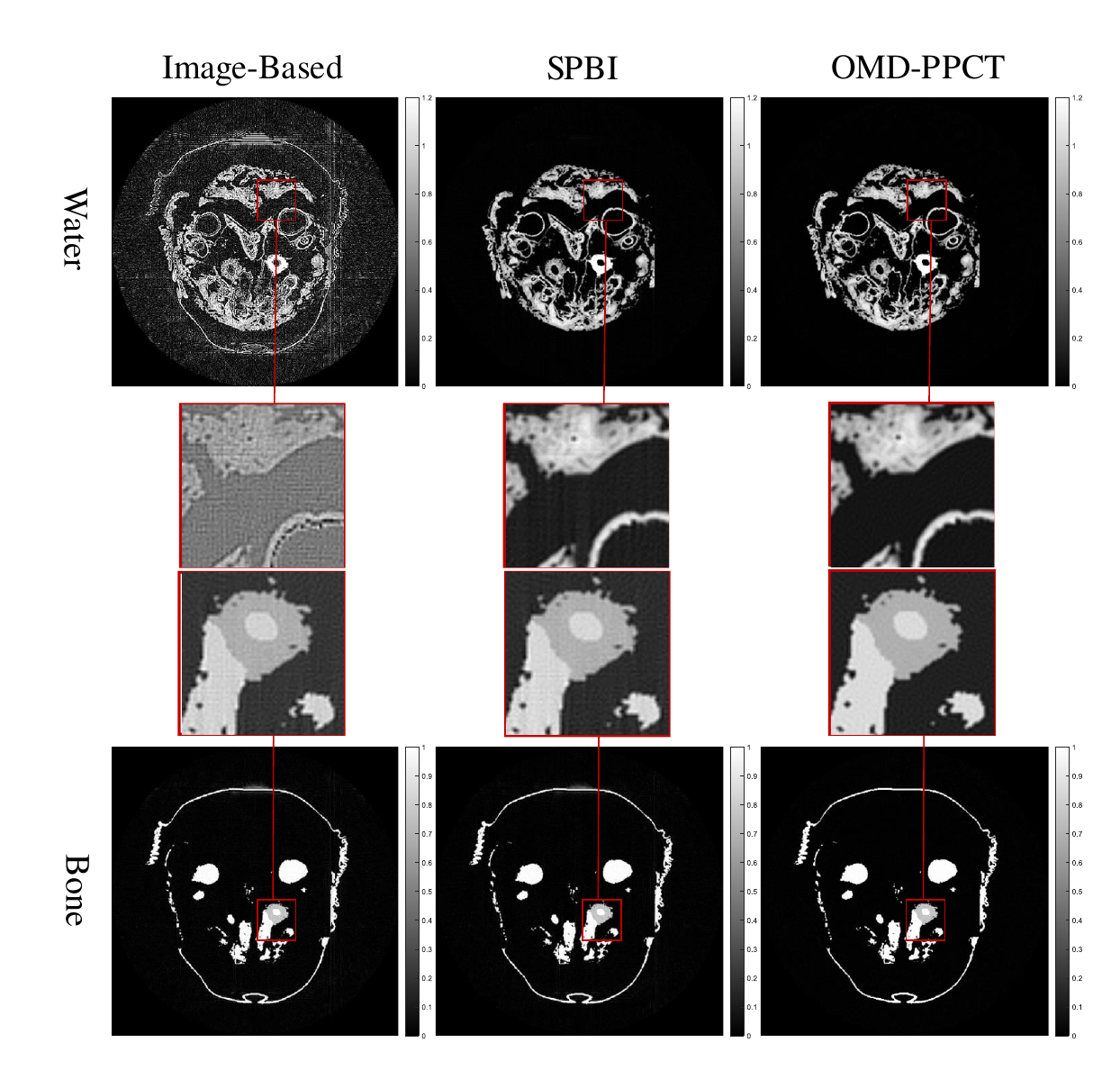}
	\caption{Material decomposition in noise-free case.}
	\label{Fig.2}
\end{figure}
\begin{figure}[]
	\centering\includegraphics[scale=0.7]{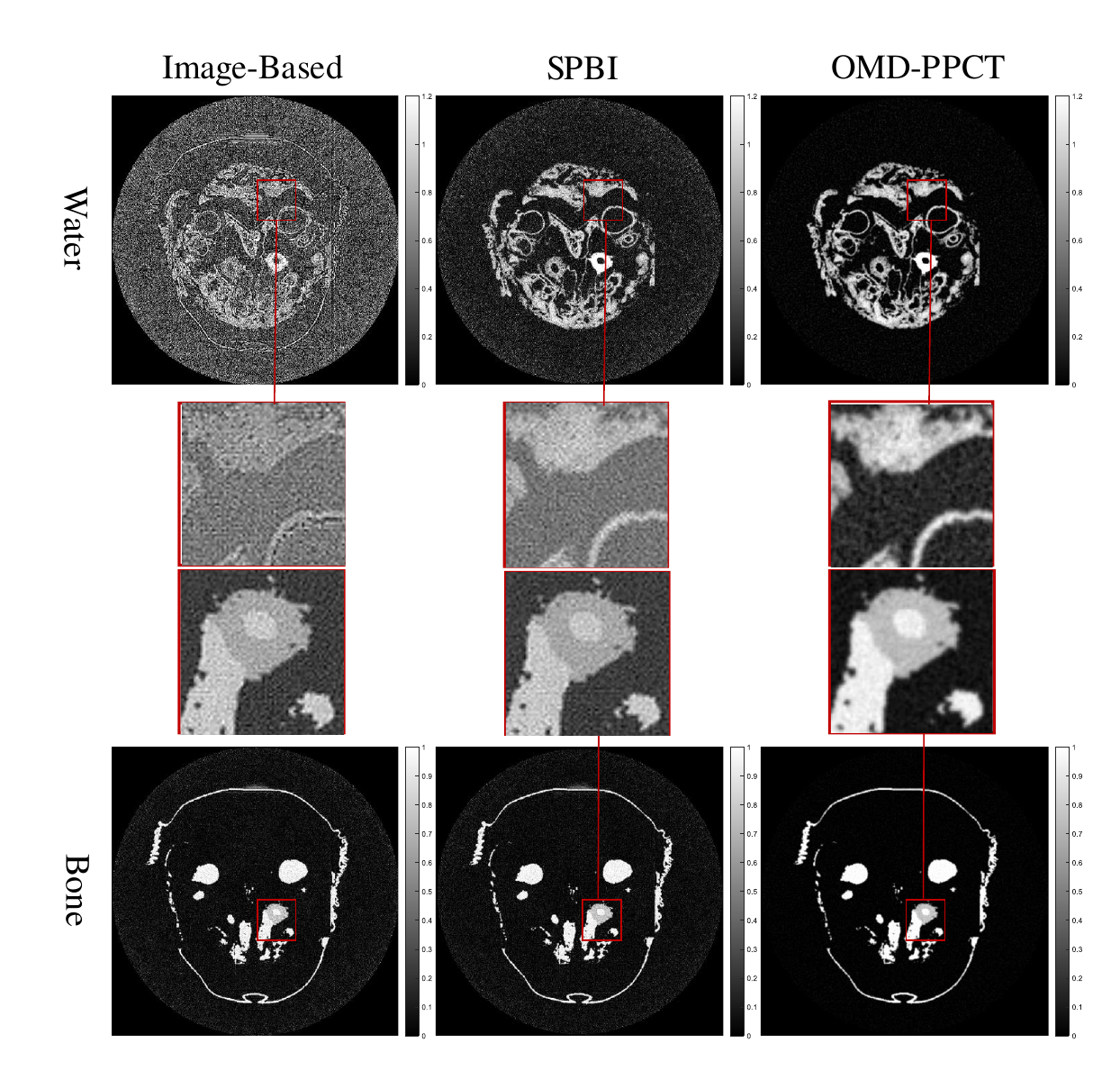}
	\caption{Material decomposition in noisy case.}
	\label{Fig.3}
\end{figure}

\begin{table}[t]
	\caption{PSNR and SSIM comparison in noise free case.}
	\begin{indented}
		\item[]\begin{tabular}{cccccccc}
			\br
			&&{Bone}&&&&{Water}\\
			\mr
            &Image-Based&SPBI&OMD-PPCT&&Image-Based&SPBI&OMD-PPCT\\
            \mr
            PSNR&20.507&21.025&$\bf{24.868}$&&5.950&21.954&$\bf{24.588}$\\

            SSIM&0.901&0.910&$\bf{0.961}$&&0.303&0.937&$\bf{0.965}$\\
			\br
		\end{tabular}
		\label{Tab.2}
	\end{indented}
\end{table}
\begin{figure}[h]
	\centering\includegraphics[scale=0.6]{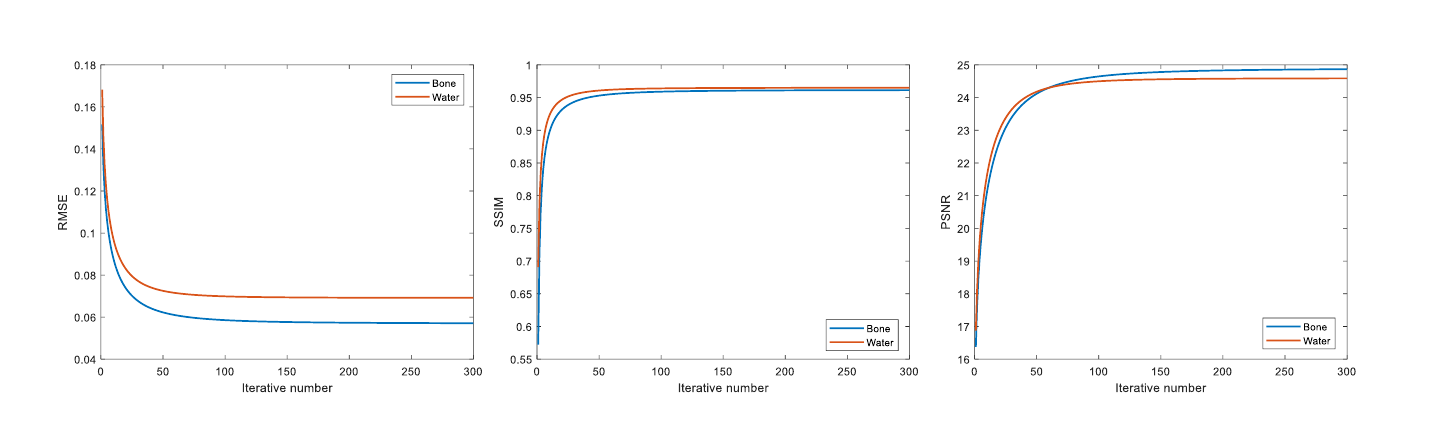}
	\caption{RMSE, PSNR and SSIM of the OMD-PPCT method.}
	\label{Fig.4}
\end{figure}

\begin{table}[t]
	\caption{PSNR and SSIM comparison in noisy case.}
	\begin{indented}
		\item[]\begin{tabular}{cccccccc}
			\br
			&&{Bone}&&&&{Water}\\
			\mr
			&Image-Based&SPBI&OMD-PPCT&&Image-Based&SPBI&OMD-PPCT\\
			\mr
			PSNR&14.750&17.533&$\bf{22.828}$&&2.267&12.772&$\bf{21.379}$\\
			
			SSIM&0.708&0.820&$\bf{0.936}$&&0.166&0.643&$\bf{0.927}$\\
			\br
		\end{tabular}
		\label{Tab.3}
	\end{indented}
\end{table}

Fig. \ref{Fig.2} and Fig. \ref{Fig.3} are the results of bone and water decomposition in noise-free case and noise case, respectively. In addition, we enlarged the details inside the red rectangle of the basic materials. The PSNR and SSIM of the basic images, as shown in Table \ref{Tab.2} and Table \ref{Tab.3}. The convergence of OMD-PPCT is proved in numerical by the curves of Root Mean Square Error (RMSE), SSIM and PSNR in Fig. \ref{Fig.4}. It is noticeable that the OMD-PPCT can effectively suppress the noise and improve the decomposed image quality. 
\section{Discussion}

In our experiments, we verify that compared to the two-step methods, the OMD-PPCT algorithm demonstrates superiority in noise suppression and material decomposition. Unlike conventional dual-energy CT, this method can effectively distinguish basic materials with weak absorption properties. Moreover, in contrast to the previous quantitative phase imaging method, this method utilizes the one-step concept and Fresnel diffraction theory to achieve material decomposition with feedback and correction from the intensity data, effectively reducing noise interference.

The energy selection is mainly determined concerning the actual imaging requirements and conditions. Noteworthy, we just focus on the reconstruction algorithm based on the imaging model in this paper. Reconstruction algorithms based on optimized models still deserve further exploration and have the potential to produce higher-quality decomposed results.
The method's drawback is that it requires some priori information about the basic material. During the experiment, multiple sets of projection data need to be acquired when using the energy integration detector, so mechanical instability may occur during the acquisition process, which leads to spending more time to align the data. However, this problem is solved when photon counting detector (PCD) is used, since the PCD has the advantage of simultaneous multi-energy acquisition \cite{Leng2019}. 

\section{Conclusion}

In this paper, we propose a one-step algorithm for material decomposition in dual-energy propagation-based phase-contrast CT. With feedback and correction in this iterative process, the proposed algorithm can obtain high-quality image from the intensity data. In addition, the Fresnel wave propagation with a Gaussian-like propagator is utilized to simulate the forward propagation, which can restrain the noise. To verify the effectiveness of the algorithm, we studied quantitative image reconstructions of sample composed by low-Z materials, which has some similarity to samples used for clinical and industrial purposes. Based on the experimental results, the proposed algorithm can effectively reduce the noise and achieve accurate decomposition results. Therefore, we believe that the algorithm has great potential for preclinical studies, especially for weakly attenuated samples.

\section*{Reference}
\bibliography{main}
 


\end{document}